\documentclass[11pt,oneside,onecolumn]{article}
\usepackage[]{fontenc}
\usepackage[dvips]{epsfig}
\usepackage{a4wide}

\makeatletter
\def\LyX{L\kern-.1667em\lower.25em\hbox{Y}\kern-.125emX\spacefactor1000}%
\newcommand{\lyxtitle}[1] {\thispagestyle{empty}
\global\@topnum\z@
\section*{\LARGE \centering \sffamily \bfseries \protect#1 }
}

\newcommand{\lyxletterstyle}{
\setlength\parskip{0.7em}
\setlength\parindent{0pt}
}

\makeatother

\lyxletterstyle
\begin{document}

\bfseries \large \hfill{}Qubit assisted Conclusive Teleportation\hfill{}
\mdseries \normalsize \bfseries \large \\
 \mdseries \normalsize 

{\hfill{}Somshubhro Bandyopadhyay\footnote{
dhom@bosemain.boseinst.ernet.in
}\hfill{} \par}

{\small \hfill{}Department of Physics, Bose Institute, 93/1 A.P.C.
Road, Calcutta -700009, India\hfill{}\hfill{}\\
\par}

{\small We present an optimal method of teleporting an unknown qubit
using any pure entangled state. \par}

\( \smallskip  \)

Quantum teleportation [1] involves, transfer of an unknown qubit from
a sender (Alice) to a receiver (Bob) via a quantum channel (a previously
shared maximally entangled pair) and using only classical communication
(for example, a phone call). However faithful teleportation [1] (and
also secure key distribution [2]) is not possible if the quantum channel
is a not a maximally entangled state. This is precisely the reason
why the entanglement concentration and purification protocols [3-8]
are of importance in quantum information theory. Recently in a very
interesting paper, Mor and Horodecki [9] obtained the optimal probability
for perfect teleportation using any pure entangled state (conclusive
teleportation). 

The present work introduces an alternative optimal method for conclusive
teleportation. The basic idea of our method is as follows. Alice first
prepares an ancilla in a state, say \( \left| \chi \right\rangle  \) (the coefficients of this state
are the Schmidt coefficients of the supplied pure entangled state)
besides her usual possession of two qubits. She now performs a joint
three particle measurement on her three qubits. It turns out that
for some of her results, Bob needs to perform only the standard rotations
\( (\sigma _{z},\sigma _{x},\sigma _{z}\sigma _{x}) \) to reconstruct the unknown state, after he gets two bits of information
from Alice. However, for any of the remaining possible set of outcomes,
Alice needs to do an optimal POVM (positive-operator value measure)
[10], which is basically a generalized measurement to discriminate
between the nonorthogonal states. She communicates her result to Bob,
who in turn rotates his qubit accordingly. The method we propose fails
sometime, but when successful the fidelity of teleportation is one.

Suppose, Alice and Bob shares a pure entangled state given by,

\begin{equation}
\label{1}
\left| \psi \right\rangle _{AB}=\alpha \left| 00\right\rangle _{AB}+\beta \left| 11\right\rangle _{AB}
\end{equation}

where we take \( \alpha ,\beta  \) to be real and \( \alpha ^{2}\geq \beta ^{2} \) (this can be assumed without any
loss of generality). 

Let the unknown state which Alice is supposed to send to Bob be,

\begin{equation}
\label{2}
\left| \phi \right\rangle _{1}=a\left| 0\right\rangle +b\left| 1\right\rangle =\left( \begin{array}{c}
a\\
b
\end{array}
\right) _{1}.
\end{equation}

Alice now prepares an ancilla qubit in the state,
\begin{equation}
\label{3}
\left| \chi \right\rangle _{2}=\alpha \left| 0\right\rangle +\beta \left| 1\right\rangle =\left( \begin{array}{c}
\alpha \\
\beta 
\end{array}
\right) _{2}.
\end{equation}
Therefore the combined
state of the four qubits is given by,\\

\begin{equation}
\label{4}
\left| \Psi \right\rangle _{12AB}=\left| \phi \right\rangle _{1}\otimes \left| \chi \right\rangle _{2}\otimes \left| \psi \right\rangle _{AB}=\left( \begin{array}{c}
a\\
b
\end{array}
\right) _{1}\otimes \left( \begin{array}{c}
\alpha \\
\beta 
\end{array}
\right) _{2}\otimes \left( \alpha \left| 00\right\rangle +\beta \left| 11\right\rangle \right) _{AB}
\end{equation}

We now observe that the state \( \left| \Psi \right\rangle _{12AB} \) can also be written as (we omit the
tensor product sign henceforth),
\[
\left| \Psi \right\rangle _{12AB}=\frac{1}{2}[(\alpha ^{2}\left| \Phi _{1}\right\rangle _{12A}+\beta ^{2}\left| \Phi _{2}\right\rangle _{12A})\left( \begin{array}{c}
a\\
b
\end{array}
\right) _{B}+(\alpha ^{2}\left| \Phi _{1}\right\rangle _{12A}-\beta ^{2}\left| \Phi _{2}\right\rangle _{12A}])\left( \begin{array}{c}
a\\
-b
\end{array}
\right) _{B}+\]

\[
(\beta ^{2}\left| \Phi _{3}\right\rangle _{12A}+\alpha ^{2}\left| \Phi _{4}\right\rangle _{12A})\left( \begin{array}{c}
b\\
a
\end{array}
\right) _{B}+(\beta ^{2}\left| \Phi _{3}\right\rangle _{12A}-\alpha ^{2}\left| \Phi _{4}\right\rangle _{12A})\left( \begin{array}{c}
-b\\
a
\end{array}
\right) _{B}]+\]

\begin{equation}
\label{5}
\frac{\alpha \beta }{\sqrt{2}}[\left| \Phi _{5}\right\rangle _{12A}\left( \begin{array}{c}
a\\
b
\end{array}
\right) _{B}+\left| \Phi _{6}\right\rangle _{12A}\left( \begin{array}{c}
a\\
-b
\end{array}
\right) _{B}+\left| \Phi _{7}\right\rangle _{12A}\left( \begin{array}{c}
b\\
a
\end{array}
\right) _{B}+\left| \Phi _{8}\right\rangle _{12A}\left( \begin{array}{c}
-b\\
a
\end{array}
\right) _{B}]
\end{equation}
\\
We now look at (5) more carefully. The most important thing to note
that, we have succeeded in writing down the combined state in a way
where one part clearly resembles the one in the BBCJPW protocol [1]
whereas the other part resembles that of Mor and Horodeckis' [9].
This in turn implies that for some of the outcomes, just standard
rotations by Bob is sufficient to construct the unknown state. If
that is not the case then of course one has to resort to POVM for
state discrimination and so on which we discuss afterwards. Now the
set \( \left\{ \Phi _{i}\right\} ,i=1,2...8 \), forms a complete orthonormal basis of the combined Hilbert
space of the three spin 1/2 particles (or two level systems) that
Alice holds and is defined by,

\[
\left| \Phi _{1}\right\rangle =\left| 000\right\rangle ;\left| \Phi _{2}\right\rangle =\left| 111\right\rangle ;\left| \Phi _{3}\right\rangle =\left| 011\right\rangle ;\left| \Phi _{4}\right\rangle =\left| 100\right\rangle \]
 
\begin{equation}
\label{6}
\left| \Phi _{5}\right\rangle =\frac{1}{\sqrt{2}}\left[ \left| 010\right\rangle +\left| 101\right\rangle \right] ;\left| \Phi _{6}\right\rangle =\frac{1}{\sqrt{2}}\left[ \left| 010\right\rangle -\left| 101\right\rangle \right] 
\end{equation}

\[
\left| \Phi _{7}\right\rangle =\frac{1}{\sqrt{2}}\left[ \left| 001\right\rangle +\left| 110\right\rangle \right] ;\left| \Phi _{8}\right\rangle =\frac{1}{\sqrt{2}}\left[ \left| 001\right\rangle -\left| 110\right\rangle \right] \]
\\
We now consider the following set of projection operators \( \left\{ P_{1},P_{2},P_{3},P_{4},P_{5},P_{6}\right\}  \)defined
by,

\[
P_{1}=P[\Phi _{1}]+P[\Phi _{2}];P_{2}=P[\Phi _{3}]+P[\Phi _{4}]\]

\begin{equation}
\label{7}
P_{3}=P[\Phi _{5}];P_{4}=P[\Phi _{6}];P_{5}=P[\Phi _{7}];P_{6}=P[\Phi _{8}]
\end{equation}

Now, in principle the measurement of an observable \( O \) is always possible
whose corresponding operator is represented by,

\begin{equation}
\label{8}
O=\sum ^{6}_{i=1}p_{i}P_{i}
\end{equation}

Eq.(8) is nothing but the spectral decomposition of the operator \( O \).
First a few words about the projection operators given by (7) is necessary.
Note that not all the projectors are of same nature. One essentially
has in the set two types of projectors, both one dimensional and two
dimensional ones. \( P_{1} \) and \( P_{2} \) are the two dimensional projectors that
projects a state onto the subspaces spanned by \( \left\{ \Phi _{1},\Phi _{2}\right\}  \) and \( \left\{ \Phi _{3},\Phi _{4}\right\}  \) respectively,
whereas the rest are all one dimensional ones.

Alice now performs a joint three particle measurement in accordance
to Eq. (8). If she obtains any one of the states belonging to the
set \( \left\{ \left| \Phi _{5}\right\rangle ,\left| \Phi _{6}\right\rangle ,\left| \Phi _{7}\right\rangle ,\left| \Phi _{8}\right\rangle \right\}  \) , each of which occurs with probability \( \frac{\alpha ^{2}\beta ^{2}}{2} \), the state of Bob's
particle is projected onto one of the following states, \( \left( \begin{array}{c}
a\\
b
\end{array}
\right)  \),\( \left( \begin{array}{c}
a\\
-b
\end{array}
\right)  \),\( \left( \begin{array}{c}
b\\
a
\end{array}
\right)  \),\( \left( \begin{array}{c}
-b\\
a
\end{array}
\right)  \). Whichever
one Bob obtains he therefore applies appropriate rotations to construct
the unknown state on his side after he receives the necessary two
bits of information from Alice. 

But Alice's measurement may also project the state onto either of
the subspaces spanned by \( \left\{ \Phi _{1},\Phi _{2}\right\}  \) and \( \left\{ \Phi _{3},\Phi _{4}\right\}  \), and each such result occurs with
probability \( \frac{\left( \alpha ^{4}+\beta ^{4}\right) }{2} \). If this is the case, then Alice needs to perform an
optimal POVM measurement (generalized measurement) in order to distinguish
between the nonorthogonal states. The explicit representation of the
operators to discriminate between two non orthogonal states can be
found in Refs. [9,10]. Suppose the result is the subspace spanned
by \( \left\{ \Phi _{1},\Phi _{2}\right\}  \). She does an optimal POVM measurement to conclusively distinguish
between the two states, \( \left( \begin{array}{c}
\alpha ^{2}\\
\beta ^{2}
\end{array}
\right) _{\{\Phi _{1};\Phi _{2}\}}\quad and\quad \left( \begin{array}{c}
\alpha ^{2}\\
-\beta ^{2}
\end{array}
\right) _{\{\Phi _{1};\Phi _{2}\}} \). Assuming (without any loss of generality)
that \( \alpha ^{2}\geq \beta ^{2} \), it turns out the probability of obtaining a conclusive result
is \( 1-\left( \frac{\alpha ^{4}-\beta ^{4}}{\alpha ^{4}+\beta ^{4}}\right) =\frac{2\beta ^{4}}{\alpha ^{4}+\beta ^{4}} \)  

Note that Alice has to inform Bob whether she succeeded or not and
that requires one bit, in addition to which she also has to send two
more bits so that Bob can perform the required rotations. 

So, given our scheme what is the probability of successful teleportation
with fidelity one? It is straightforward to obtain that the probability
\em p \em of having perfect teleportation is 
\begin{equation}
\label{9}
p=2\beta ^{4}+2\alpha ^{2}\beta ^{2}=2\beta ^{2}
\end{equation}

Comparing our probability of perfect teleportation with that of Mor
and Horodecki [6] (\( p_{MH} \)) which is \( 2\beta ^{2} \), we see that \( p=p_{MH} \). So our result also
brings about the optimal probability of perfect teleportation with
any pure entangled state.

In summary, we have described an optimal method for teleporting an
unknown quantum state using any pure entangled state. A positive implication
of our strategy is in its partial dependence on POVM to achieve perfect
teleportation. We have seen that for some of Alice's outcomes, only
standard rotations need to be performed by Bob to get the unknown
state . Nevertheless the cost one has to pay for it is a joint three
particle measurement. 

I wish to acknowledge Guruprasad Kar and Anirban Roy for many stimulating
discussions. I'm grateful to Anthony Chefles and H. K. Lo for pointing
out an error in the earlier version of this work to my attention.

\( \smallskip  \)

[1] C. H. Bennett, G. Brassard, C. Crepeau, R. Jozsa, A. Peres and
W. K. Wootters, Phys. Rev. Lett. \bfseries 70\mdseries , 1895
(1993).

[2] A. K. Ekert, Phys. Rev. Lett. \bfseries 67\mdseries , 661
(1991).

[3] C. H. Bennett, H. J. Bernstein, S. Popescu, and B. Schumacher,
Phys. Rev. A \bfseries 53\mdseries , 2046 (1996).

[4] S. Bose, V. Vedral and P. L Knight, quant-ph/9812013.

[5] H. K. Lo and S. Popescu, quant-ph/9707038.

[6] C. H. Bennett, D. P. DiVincenzo, J. A. Smolin and W. K. Wootters,
Phys. Rev. A 54, 3824 (1996).

[7] C. H. Bennett, G. Brassard, S. Popescu, B. Schumacher, J. A. Smolin
and W. K. Wootters, Phys. Rev. Lett. \bfseries 76\mdseries ,
722 (1996). 

[8] D. Deutsch, A. Ekert, R. Jozsa, C. Macchiavello, S. Popescu and
A. Sanpera, Phys. Rev. Lett. \bfseries 77\mdseries , 2818 (1996).

[9] T. Mor and P. Horodecki, quant-ph/9906039. 

[10] A. Peres, Quantum Theory: Concepts and Methods (Kluwer, Dordrecht,
1993).

\end{document}